\documentclass{sf2a-conf2014}
\usepackage{graphicx}
\usepackage{hyperref}
\usepackage[]{natbib}  
\usepackage{amssymb,amsmath}

\def\BibTeX{{\rm B\kern-.05em{\sc i\kern-.025em b}\kern-.08em
    T\kern-.1667em\lower.7ex\hbox{E}\kern-.125emX}}
\bibpunct{(}{)}{;}{a}{}{,}  

\newcommand\del{\bb{\nabla}}

\newcommand{\BV}{Brunt-V\"ais\"al\"a\ }

\newcommand\bb[1] {   \mbox{\boldmath{$#1$}}  }

\begin{document}

\TitreGlobal{SF2A 2014}

\title{Gravito-inertial modes in a differentially rotating spherical shell}

\runningtitle{Gravito-inertial modes in a differentially rotating spherical shell}

\author{G. M. Mirouh$^{1,}$}\address{Universit\'e de Toulouse, UPS-OMP, IRAP, Toulouse, France}
\address{CNRS, IRAP, 14 avenue Edouard Belin, 31400 Toulouse, France}

\author{C. Baruteau$^{1,2}$}
\author{M. Rieutord$^{1,2}$}
\author{J. Ballot$^{1,2}$}

\setcounter{page}{237}


\maketitle


\begin{abstract}
Oscillations have been detected in a variety of stars, including intermediate- and high-mass main sequence stars. 
While many of these stars are rapidly and differentially rotating, the effects of rotation on oscillation modes are poorly known. 
In this communication we present a first study on axisymmetric gravito-inertial modes in the radiative zone of a differentially rotating star. 
These modes probe the deep layers of the star around its convective core. 
We consider a simplified model where the radiative zone of a star is a linearly stratified rotating fluid within a spherical shell, 
with differential rotation due to baroclinic effects. 
We solve the eigenvalue problem with high-resolution spectral simulations and determine the propagation domain of the waves through 
the theory of characteristics. We explore the propagation properties of two kinds of modes: 
those that can propagate in the entire shell and those that are restricted to a subdomain. 
Some of the modes that we find concentrate kinetic energy around short-period shear layers known as attractors. 
We characterise these attractors by the dependence of their Lyapunov exponent with the \BV frequency of the background and the oscillation frequency of the mode. 
Finally, we note that, as modes associated with short-period attractors form dissipative structures, 
they could play an important role for tidal interactions but should be dismissed in the interpretation of observed oscillation frequencies. 
\end{abstract}



\section{Introduction}
High-precision photometry provided by space missions, such as CoRoT 
and Kepler, 
have allowed measuring many oscillation frequencies 
in a variety of stars. In rotating stars, low-frequency modes correspond to modes restored by buoyancy and Coriolis forces, 
and are called gravito-inertial modes.
Their excitation can be provided by internal mechanisms, such as the $\kappa$-mechanism. 
For planet-harbouring stars, tidal effects may excite such modes. 
Determining how tidally-excited waves deposit their energy and angular momentum helps predicting the orbital evolution of close-in planets.

We wish to have more insights into the modal properties of high- and intermediate-mass main sequence stars. 
High rotation rates have been detected in most of these stars \citep{royer09}, while \citet{ELR13} 
showed that the radial differential rotation increases throughout the evolution of these stars.
The oscillation properties of rapidly and differentially rotating stars are less constrained than those of slow rotators. For instance,
their spectra show no regular patterns as shown in delta Scuti stars \citep{mirouh_etal14}.
The fundamental parameters and the mode amplitudes are also modified by rotation, as shown for fast-rotating SPB stars 
detected outside of their expected instability domain \citep{mow13,salmon14}.

Inertial modes in a differentially rotating convective layer have been studied by \citet{BR13}, 
but the influence of differential rotation on the properties of gravito-inertial modes in a radiative zone is still an uncharted territory.
In those stars, gravito-inertial modes probe the internal layers of the radiative zone, around the convective core. 
A better characterisation of the oscillations could help constrain the physical model, in particular the core size.

\section{Model}
\label{sec:model}
We model the outer radiative region of a star as a viscous fluid enclosed in a spherical shell. The convective core is not modeled. 
We impose a linear background temperature gradient through the radiative shell ($\del T = -\beta\bb{r} / R$, 
where $T$ is the background temperature, and $\beta$ a positive constant). 
This gradient yields a linear stable stratification \citep{chandra61}, where the \BV frequency reads $n(r) = N \times r$ with $N$ a constant.

\citet{R06} has shown that the baroclinic flow (i.e. the flow resulting from the combined effects of rotation and stratification) gives,
for no-slip boundary conditions on both sides of the shell \citep{HR14}, the following angular velocity profile $\Omega(r)$:
\begin{equation*}
  \frac{\Omega(r)}{\Omega(R)} = 1 + \int\limits^R_r{\frac{n^2(r')}{r'}} dr' = 1 + \frac{N^2}{2} \left(1-\frac{r^2}{R^2}\right).
\end{equation*}
This is the configuration that we have adopted, resulting in a shellular differential rotation. We also make use of the Boussinesq approximation.

We determine the properties of the linear oscillations in this model using two methods \citep[see e.g.][]{RV97, DRV99, BR13}. 
\begin{enumerate}
\item 
  We solve the eigenvalue problem of the stellar oscillations with a spectral code. 
  This is done by considering the linearised equations of motion, energy, and mass conservation, in which our quantities are decomposed as the sum 
  of a background state and a perturbation in $\exp(i \omega t)$ (with $\omega$ the mode frequency in the inertial frame). 
  The equations are then projected onto the spherical harmonics, as described in \citet{R87}.
  We compute eigenvalues (and the associated eigenmodes) of the fully dissipative system.
  
\item
  We compute the paths of characteristics, in the non-dissipative limit.
  This approach consists in neglecting the viscous and thermal dissipations and allows us to reduce the set of linearised equations
  as a partial differential equation for the pressure perturbation, which is often referred to as the pressure operator. 
  From this operator of hyperbolic or mixed type, we extract the equation of the characteristics
  \citep{FS82a}. 
\end{enumerate}

In the present study the dissipative properties of the fluid are characterised by the Prandtl and the Ekman numbers, which are respectively defined as:
\begin{equation}
  {\rm Pr} = \frac{\nu}{\kappa},\qquad {\rm E} = \frac{\nu}{\Omega(R) R^2},
\end{equation}
where $\nu$ is the kinematic viscosity and $\kappa$ the thermal diffusivity of the fluid.

In stars, $\rm{Pr} \sim 10^{-5}$ and $\rm{E} \sim 10^{-10} - 10^{-12}$. However, such small values require considerable spatial resolution, 
making the numerical problem ill-conditioned \citep{VRBF07}. 
For a first numerical exploration, we set $\rm{Pr} = 10^{-2}$ and $\rm{E} = 10^{-8}$ and scan the $(N^2, \omega)$ plane.

\section{Mode classification}
The pressure operator defined in section \ref{sec:model} is of mixed type : depending on the parameters and the position in the star, 
the solutions may be evanescent or oscillatory. 
Therefore, for a given set of $(N^2, \omega)$, eigenmodes may occupy only a fraction 
of the spherical shell. This permits a classification of the modes in two categories, as in \citet{BR13}. 

If waves can propagate in the whole shell, we call these modes D modes (for Differential rotation). 
Conversely, if the operator is elliptic in the whole shell, the oscillations are damped and we see no mode. 
If the type of the operator changes in the shell, the waves are confined to the hyperbolic domain in the shell. 
The boundaries of the propagation domain are called critical or turning surfaces,
and we call DT modes the modes exhibiting this behaviour (for Differential rotation with a Turning surface). 

\begin{figure}[h!]
  \includegraphics[width=0.3\textwidth]{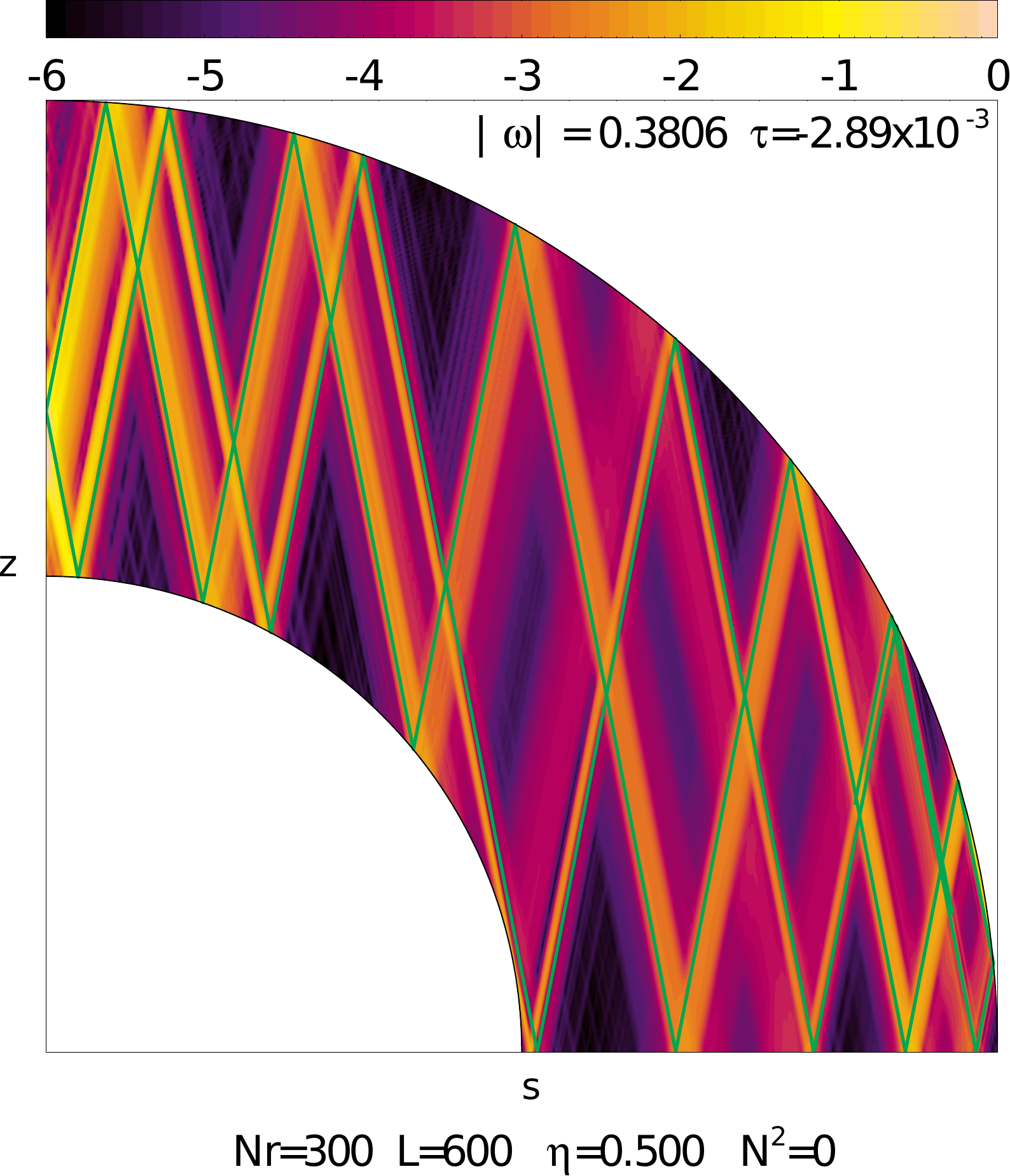} \hfill 
  \includegraphics[width=0.3\textwidth]{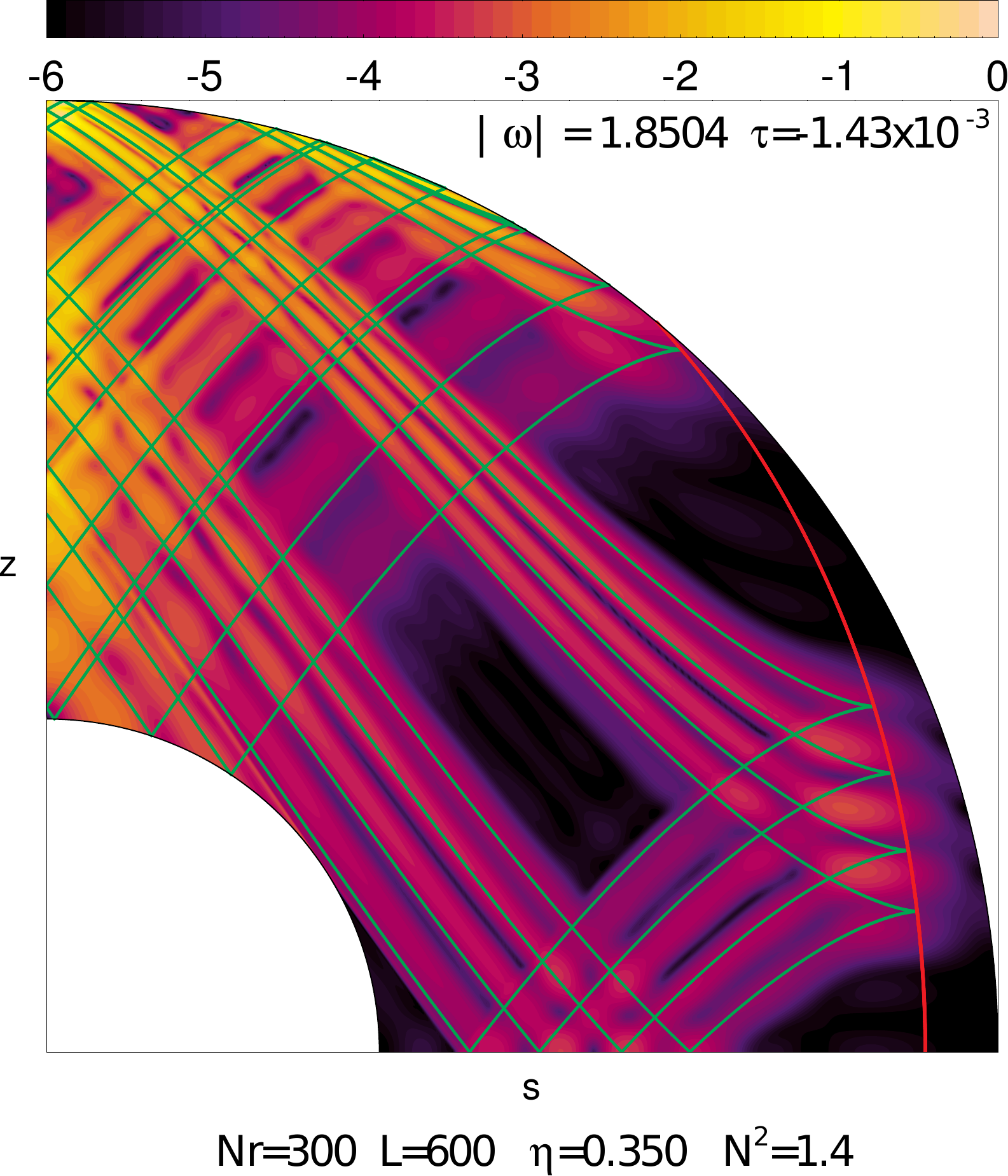}  \hfill
  \includegraphics[width=0.3\textwidth]{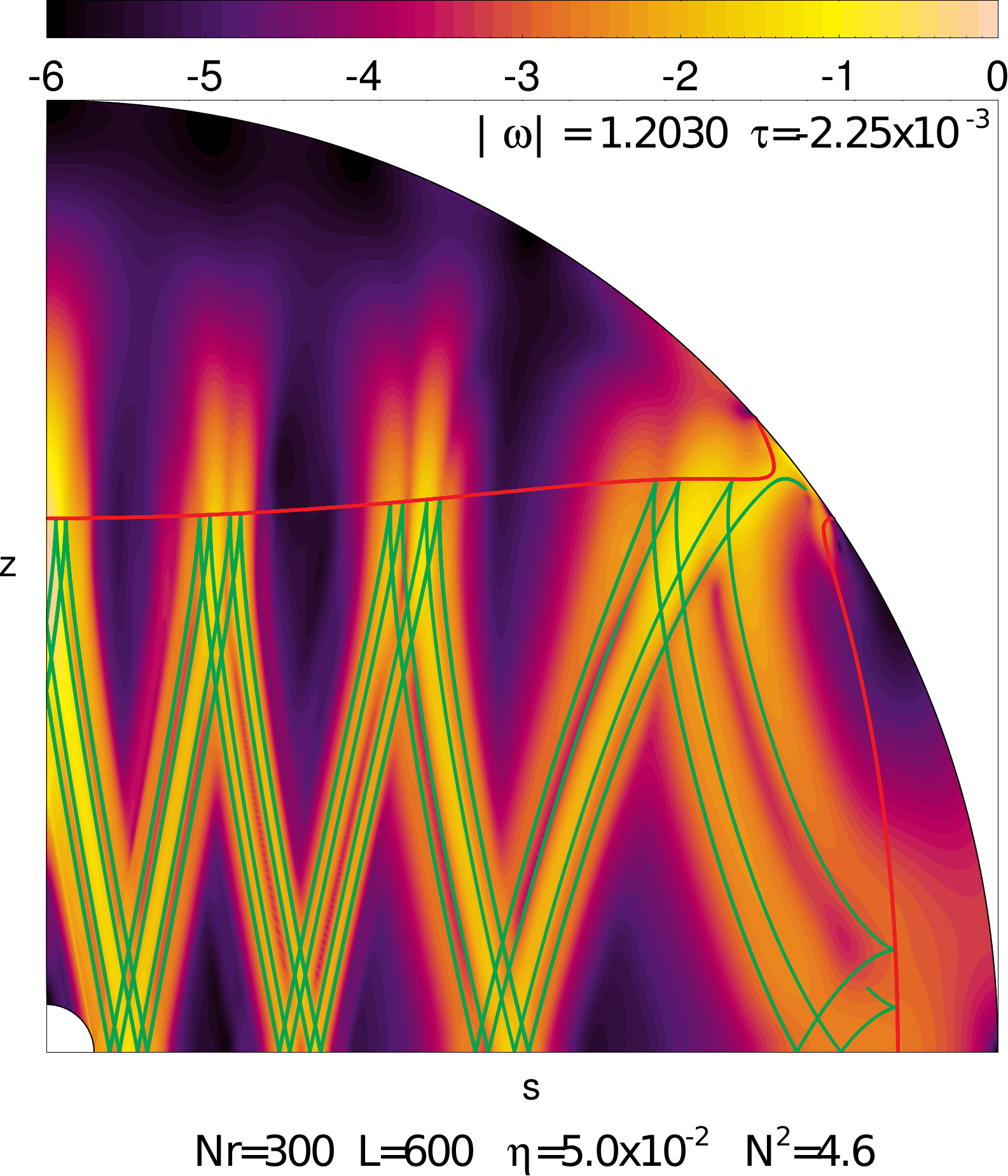}  
  \caption{Meridional slices of kinetic energy obtained by solving the dissipative linearised hydrodynamics equations, 
  with attractors (green) and turning surfaces (red) overplotted. The energy is plotted on a logarithmic scale and normalised to its maximum value.}
  \label{fig:modes}
\end{figure}

Figure \ref{fig:modes} shows the kinetic energy distribution of some axisymmetric ($m=0$) modes for various values of
$\eta$, $N^2$ and $\omega$, in a meridional plane.
The energy is focused along shear layers, which correspond to the trajectory of characteristics computed in the non-diffusive case.
Varying parameters E and Pr changes the focusing of the modes towards the characteristics. At vanishing viscosities, the shear layers become singular. 
The mode on the left is an inertial mode with solid-body rotation ($N^2=0$, $\Omega$ constant), and is a D mode spanning the whole shell. The two other 
modes are gravito-inertial modes in a differentially rotating background. The oscillation is confined to a subdomain, bounded by one or several turning surfaces computed
analytically.

\begin{figure}[h!]
  \hspace{0.1\textwidth} \includegraphics[angle=-90,width=0.8\textwidth]{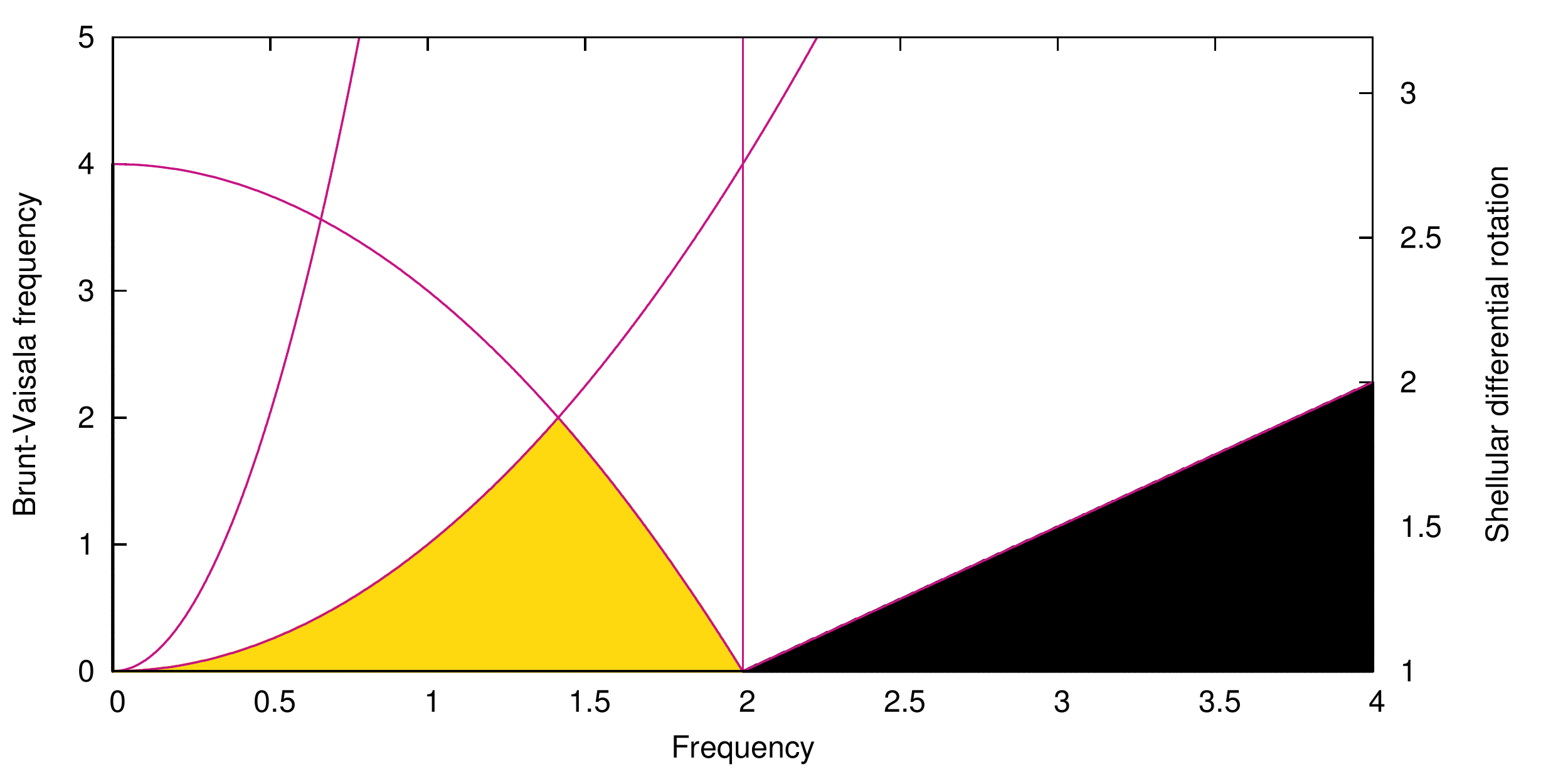} 
  \caption{Map of the axisymmetric D modes (yellow) and DT modes (white), for $\eta=0.35$. 
    The solid lines correspond to changes in the mode geometry shown in the miniatures. 
    No modes exist in the black domain.}
  \label{fig:map}
\end{figure}

We study the propagation properties of the oscillations for a range of $N^2$ and $\omega$.
Investigating the mathematical properties of the pressure operator yields the curves separating the D and DT modes geometries. 
Figure \ref{fig:map}$\ $summarises the various classes of modes in the parameter space $(N^2, \omega)$, each domain corresponding to a different geometry. 
The boundaries between the domains are calculated analytically, and confirmed by computing the type of the pressure operator numerically.

\section{Attractors and dissipation}
The modes shown in figure \ref{fig:modes} feature only modes associated with short-period attractors. 

Such modes are expected to be very damped, and thus should be excluded when trying to identify observed frequencies. 
However, as their dissipation rate is high, these modes are important for tidal interaction.

\begin{figure}[h!]
  \hspace{0.2\textwidth} \includegraphics[width=0.6\textwidth]{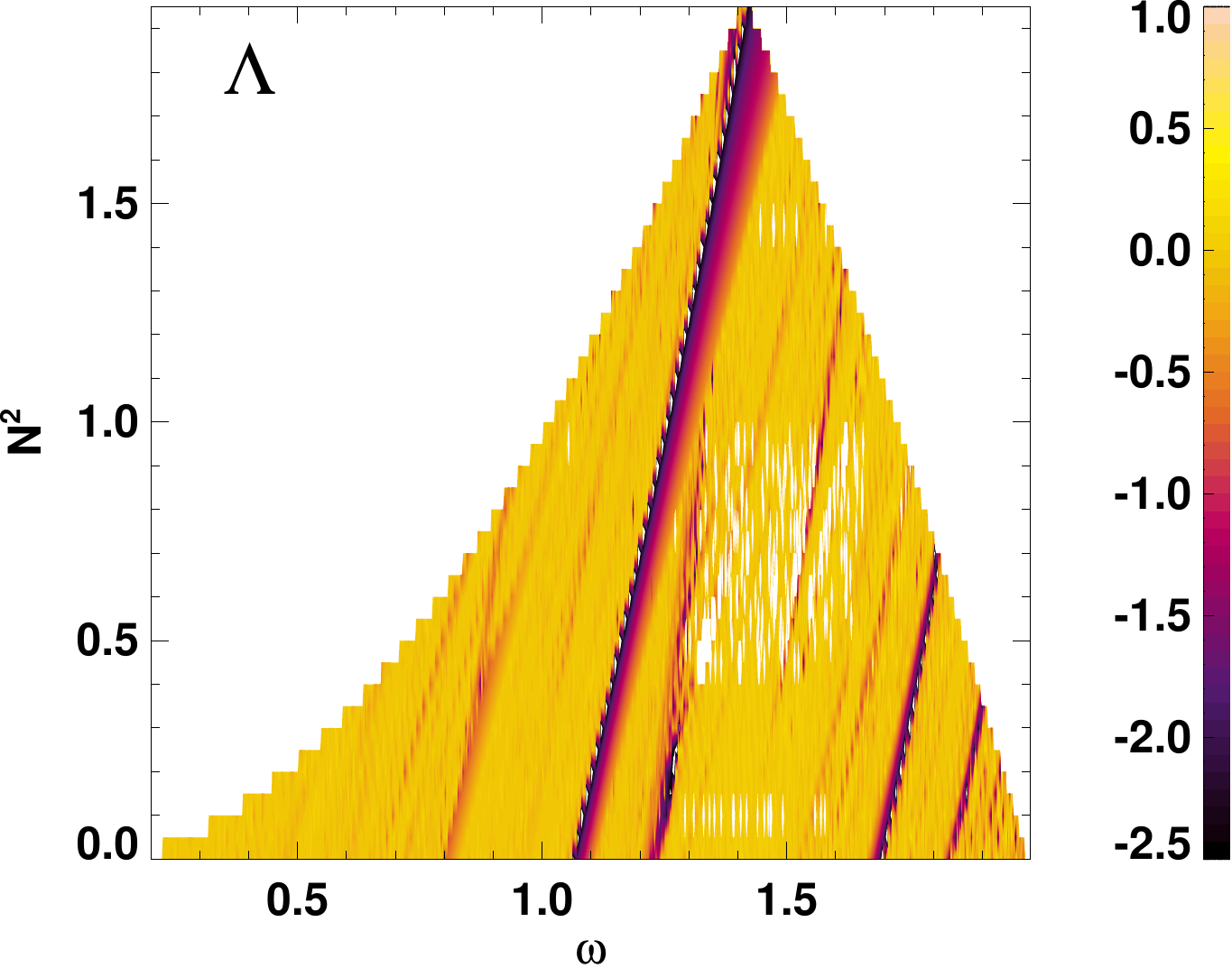} 
  \caption{Lyapunov exponent for D modes, as a function of stratification and wave frequency.}
  \label{fig:Lyap}
\end{figure}

To determine whether characteristics tend or not towards a periodic structure, we compute the Lyapunov exponent of the characteristic trajectories
\citep[e.g.][]{RGV01}. 
This exponent quantifies the convergence of the characteristics towards attractors. 
We have computed the Lyapunov exponent for all the D modes, as shown in figure \ref{fig:Lyap}. We see ridges where the Lyapunov exponent is very negative,
corresponding to modes featuring short-period attractors, which slightly change their frequency and geometry when $N^2$ varies. However, most of the domain corresponds to
longer-period attractors compatible with low dissipation, and therefore with excitable stellar pulsations.

\section{Conclusions and future prospects}
For the first time, we compute the oscillations of a differentially rotating radiative region of star, where differential rotation is part of the 
baroclinic flow triggered by the combined effects of rotation and stable stratification. 

We have given a first view of axisymmetric eigenmodes which may propagate in such a background flow. The next steps include investigating non-axisymmetric modes
and using more realistic \BV frequency profiles, 
before tackling the more realistic configuration of a two-dimensional compressible stellar model, as in the ESTER models \citep{ELR13}.

\bibliographystyle{aa}  

\begin{thebibliography}{14}
    \expandafter\ifx\csname natexlab\endcsname\relax\def\natexlab#1{#1}\fi

  \bibitem[{{Baruteau} \& {Rieutord}(2013)}]{BR13}
        {Baruteau}, C. \& {Rieutord}, M. 2013, Journal of Fluid Mechanics, 719, 47

  \bibitem[{Chandrasekhar(1961)}]{chandra61}
    Chandrasekhar, S. 1961, Hydrodynamic and hydromagnetic stability (Clarendon
      Press, Oxford)

  \bibitem[{Dintrans {et~al.}(1999)Dintrans, Rieutord, \& Valdettaro}]{DRV99}
         Dintrans, B., Rieutord, M., \& Valdettaro, L. 1999, J. Fluid Mech., 398, 271

  \bibitem[{{Espinosa Lara} \& {Rieutord}(2013)}]{ELR13}
        {Espinosa Lara}, F. \& {Rieutord}, M. 2013, A\&A, 552, A35

  \bibitem[{{Friedlander} \& {Siegmann}(1982a)}]{FS82a}
       {Friedlander}, S. \& {Siegmann}, W. 1982a, J. Fluid Mech., 114, 123

  \bibitem[{Hypolite \& Rieutord(2014)}]{HR14}
                Hypolite, D. \& Rieutord, M. 2014, to appear in A\&A, 1, 1

  \bibitem[{{Mirouh} {et~al.}(2014){Mirouh}, {Reese}, {Espinosa Lara}, {Ballot}, \&
        {Rieutord}}]{mirouh_etal14}
        {Mirouh}, G.~M., {Reese}, D.~R., {Espinosa Lara}, F., {Ballot}, J., \& {Rieutord}, M.
          2014, in IAU Symposium, Vol. 301, IAU Symposium, ed. J.~A. {Guzik}, W.~J.
            {Chaplin}, G.~{Handler}, \& A.~{Pigulski}, 455--456

  \bibitem[{{Mowlavi} {et~al.}(2013){Mowlavi}, {Barblan}, {Saesen}, \&
            {Eyer}}]{mow13}
            {Mowlavi}, N., {Barblan}, F., {Saesen}, S., \& {Eyer}, L. 2013, \aap, 554, A108

  \bibitem[{Rieutord(1987)}]{R87}
            Rieutord, M. 1987, Geophys. Astrophys. Fluid Dyn., 39, 163

  \bibitem[{Rieutord(2006)}]{R06}
            Rieutord, M. 2006, A\&A, 451, 1025

  \bibitem[{Rieutord {et~al.}(2001)Rieutord, Georgeot, \& Valdettaro}]{RGV01}
                                   Rieutord, M., Georgeot, B., \& Valdettaro, L. 2001, J. Fluid Mech., 435, 103

  \bibitem[{Rieutord \& Valdettaro(1997)}]{RV97} Rieutord, M. \& Valdettaro, L. 1997, J. Fluid Mech., 341, 77

  \bibitem[{{Royer}(2009)}]{royer09}
            {Royer}, F. 2009, in Lecture Notes in Physics, Berlin Springer Verlag, Vol.
              765, The Rotation of Sun and Stars, ed. J.-P. {Rozelot} \& C.~{Neiner},
              207--230

  \bibitem[{{Salmon} {et~al.}(2014){Salmon}, {Montalb{\'a}n}, {Reese}, {Dupret},
                   \& {Eggenberger}}]{salmon14}
                  {Salmon}, S.~J.~A.~J., {Montalb{\'a}n}, J., {Reese}, D.~R., {Dupret}, M.-A., \&
                    {Eggenberger}, P. 2014, \aap, 569, A18

  \bibitem[{Valdettaro {et~al.}(2007)Valdettaro, Rieutord, Braconnier, \&
                      Fraysse}]{VRBF07}
                      Valdettaro, L., Rieutord, M., Braconnier, T., \& Fraysse, V. 2007, J. Comput.
                        and Applied Math., 205, 382

\end{thebibliography}

\end{document}